# Using Brain Connectivity Measure of EEG Synchrostates for Discriminating Typical and Autism Spectrum Disorder

Wasifa Jamal, Saptarshi Das, Koushik Maharatna, *Member, IEEE*, Doga Kuyucu, Federico Sicca, Lucia Billeci, Fabio Apicella, and Filippo Muratori

*Abstract*—In this paper we utilized the concept of stable phase synchronization topography – synchrostates – over the scalp derived from EEG recording for formulating brain connectivity network in Autism Spectrum Disorder (ASD) and typically-growing children. A synchronization index is adapted for forming the edges of the connectivity graph capturing the stability of each of the synchrostates. Such network is formed for 11 ASD and 12 control group children. Comparative analyses of these networks using graph theoretic measures show that children with autism have a different modularity of such networks from typical children. This result could pave the way to a new modality for possible identification of ASD from non-invasively recorded EEG data.

## I. Introduction

Recent researches have established phase synchronization to be a biological mechanism of communication between brain regions [1]. Studies have found evidences of short- and long range phase synchrony as a key manifestation of information integration process in brain during cognition [2] and its study provides an independent dimension of understanding information processing in the brain [3].

ASD is a life-long condition that is characterized by lack of empathy and atypical behavior and early intervention is projected as the best possible way for managing it. Prominent research conducted in the field of ASD suggests that autism is caused by deficit of neural level information integration due to under-functioning integrative circuitry [4, 5]. The functional Magnetic Resonance Imaging (fMRI) based studies have reported evidences of overall functional under-connectivity in autism compared to controls [6]. Tommerdahl *et al.* [7], in their study of sensory perception in autism detected local under-connectivity in autistic adults. Therefore quantitative characterization of the connectivity derived from phase synchronization characteristics in ASD patients may lead to an effective diagnostic modality enabling intervention at appropriate stage.

The existence of unique set of millisecond-order stable phase synchronized topographies over the scalp from EEG recording (termed *synchrostate*) during execution of a face perception task in an adult was first demonstrated by Jamal *et al.* [8] showing that the optimal number of synchrostates was consistently in a small range (3–6) over a number of repeated trials and the inter-synchrostate switching follows a well-behaved temporal sequence during the task. It also showed that some of these synchrostates occur more frequently compared to others during the execution of the task. The present study extends this concept to capture the stability of the individual synchrostates with a synchronization index which is representative of information exchange mechanism within the brain. We show that when this synchronization index is used to build a connectivity map over the scalp corresponding to the maximum or minimum occurring synchrostates, its analysis using graph-theoretic measures effectively distinguishes between children with ASD and age matched controls when presented with faces with different emotions *viz.* fearful, happy and neutral, as studied in Apicella *et al.* [9], leading to a possible new way of diagnosing ASD children from EEG analysis.

Complex network measures have been used in recent studies to quantify brain networks [10]. Bellmore and Sporns [11] demonstrated the usefulness of complex network approaches to study anatomical and functional brain networks. Specifically, modularity is a sophisticated measure that quantitatively characterizes the segregation property of a network reflecting the degree to which a network can be subdivided into a group of nodes with small number of between-group links (edges) and large number of within-group links [12]. Since measure of segregation in a brain network quantifies its ability for specialized processing and therefore describes the organizational property of the local connectivity during information integration, we deem that modularity could be a useful index for characterizing an ASD brain. Thus in this paper, we use modularity to quantitatively compare network topographical differences between two age-matched populations *viz.* typical and ASD and show that it may be possible to distinguish these two groups from modularity measure. The rest of the paper is structured as follows: Section II gives a theoretical background of synchrostate formulation and modularity measure along with the experimental protocol, Section III analyses the experimental results and establishes the possibility for distinguishing ASD group from age-matched controls and the conclusions are drawn in Section IV.

* The work presented in this paper was supported by FP7 EU funded MICHELANGELO project, Grant Agreement # 288241.

W. Jamal, S. Das, D. Kuyucu, K. Maharatna are with the School of Electronics and Computer Science, University of Southampton, Southampton SO17 1BJ, United Kingdom (e-mail: {wj4g08, sd2a11, km3, dk2g09}@ecs.soton.ac.uk).

F. Sicca, F. Apicella, F. Muratori are with Stella Maris Scientific Institute, Viale del Tirreno, 331, I-56018, Calambrone, Pisa, Italy (email: {fsicca, fapicella, filippo.muratori}@inpe.unipi.it).

L. Billeci is with the Institute of Clinical Physiology (IFC), National Council of Research (CNR), Pisa, Italy (email: lucia.billeci@ifc.cnr.it).

## II. THEORETICAL BACKGROUND AND EXPERIMENTS

### A. Mathematical Foundation of Synchrostate

Complex Morlet wavelet transform when applied on EEG signal yields instantaneous amplitude and phase components of the signal captured at an EEG electrode. This instantaneous phase information is used to construct a phase difference time series relative to each frequency or wavelet scale. Unsupervised *k*-means clustering [13] is applied to these difference matrices to form compact clusters or states during which there is little variation in phase topography and hence have been termed as synchrostates [8] and inter-synchrostates switching follows a temporal sequence responsible to complete that specific task. However since the clustering technique only identified the synchrostates but does not capture their temporal stability period, for its quantitative estimation we use the phase synchronization index [14] given in (1) which is an inverse statistical analogue of variance.

$$\Gamma_{xy}(a) = \frac{1}{N}\sqrt{\left[\sum_t \cos(\Delta\varphi_{xy}(a,t))\right]^2 + \left[\sum_t \sin(\Delta\varphi_{xy}(a,t))\right]^2} \quad (1)$$

$$\Delta\varphi_{xy}(a,t) = \varphi_x(a,t) - \varphi_y(a,t)$$

Here, $\varphi_x(a,t)$ and $\varphi_y(a,t)$ are the arguments of the complex wavelet transform in terms of time *t* and wavelet scale *a* (function of frequency) of signals $x(t)$ and $y(t)$ respectively. Also, $N$ is the number of data points in the time series and $\Gamma_{xy}(a) \in [0,1]$. The synchronization index thus calculated between all pairs of EEG electrodes for each synchrostate can be used to form a corresponding connectivity matrix where its high values indicate a high degree of synchronization between two channels and hence higher degree of connectivity.

Now the brain connectivity graph can be structured from the connectivity matrix formulated in (1). In order to quantitatively characterize the connectivity graph, as mentioned in Section I, we use modularity as a measure of segregation and therefore the ability of specialized processing in the brain networks reflecting the degree of local connectivity during specialized processing [15]. The modularity of a complex weighted network graph can be expressed as [12]:

$$Q^w = \frac{1}{l^w}\sum_{i,j\in N}\left[w_{ij} - \frac{k_i^w k_j^w}{l^w}\right]\delta_{m_i,m_j} \quad (2)$$

where, $w_{ij}$ is the connection weights, $k_i^w = \sum_{j\in N} w_{ij}$ is the weighted degree, $l^w = \sum_{i,j\in N} w_{ij}$ is the sum of all weights in the network. Also, $\delta_{m_i,m_j} = 1$ if $m_i = m_j$, and 0 otherwise.

### B. Experimental Paradigm with Typical and ASD Children

The experimental population for this study consists of 12 healthy control children with neuro-typical development and 11 children diagnosed as ASD – both groups in the age range of 6 – 13 years [9]. Both groups were presented three types of emotional face stimuli with standardized face expressions of happiness, fear and neutral expression which were presented 40 times [9]. 128-channel EEG was used to continuously record data at 250 samples per second. Data was segmented into 1000 ms epoch with 150 ms of baseline and 850 ms of post stimuli response. Epochs over a threshold of 200μV were rejected as artifacts. Data was baseline corrected and band pass filtered from 0.5 Hz to 50 Hz to remove drift and noise.

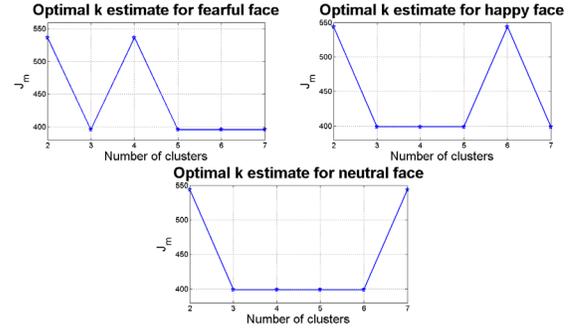

Figure 1. Optimum *k* estimate for gamma band for typical children.

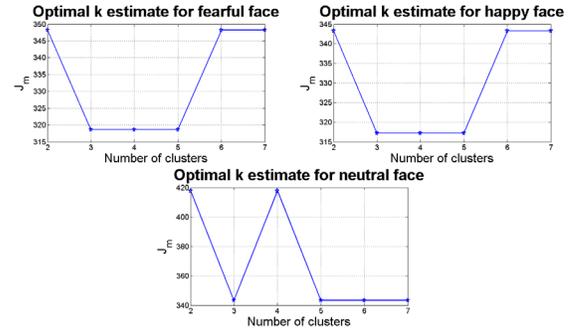

Figure 2. Optimum *k* estimate for gamma band for children with ASD.

With the EEG data thus processed, we followed the procedures described in [8] to form the synchrostates. However, since we are interested in the group results we averaged the phase response of all subjects in each class - ASD and typical – before we ran the clustering algorithm to determine the optimal number of synchrostates. We only focus on the gamma band (30Hz - 48 Hz) response of both ASD and typical children as previous research confirms modulations in the gamma band activity induced during visual information processing [16]. The clustering results show that for each case of ASD and typical, the number of synchrostates was consistently three (determined by observing a significant knee in the cost function in Figs. 1-2) for each of the three stimuli.

## III. VISUALIZATION AND ANALYSIS OF THE RESULTS

### A. Synchrostates in Typical and ASD Children

Once formulated, the phase topographical plots for the synchrostates corresponding to each of the stimuli show marked comparative difference between the typical control

(Fig. 3 – 5) and ASD group (Fig. 6 – 8). Although the phase topography for the typical group remains more or less the same for the three stimuli, the configurations are visibly different in the ASD group under different stimuli implying different network configurations for the selected stimuli compared to the typical group.

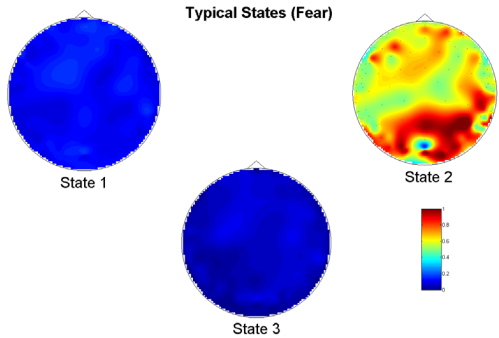

Figure 3. Synchrostates for typical children with fearful face stimulus.

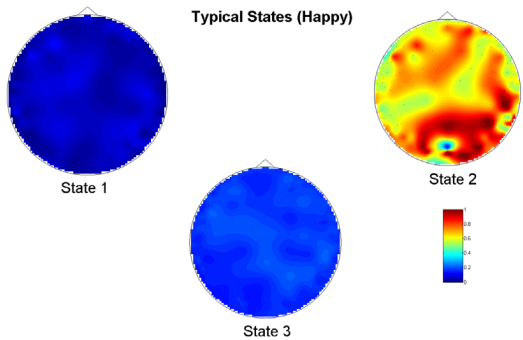

Figure 4. Synchrostates for typical children with happy face stimulus.

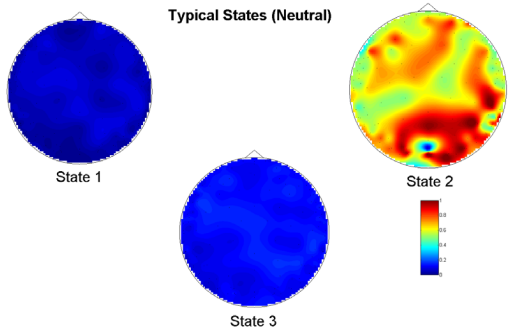

Figure 5. Synchrostates for typical children with neutral face stimulus.

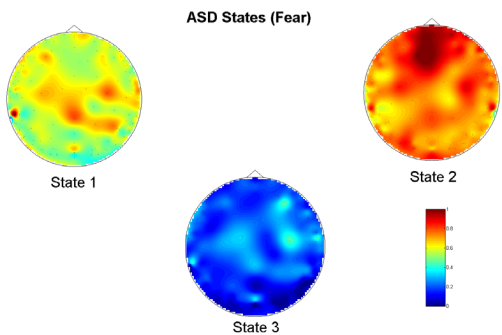

Figure 6. Synchrostates for ASD children with fearful face stimulus.

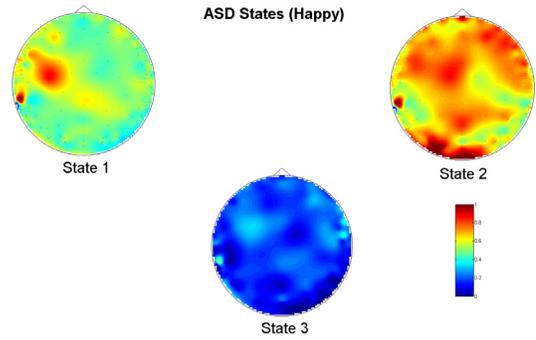

Figure 7. Synchrostates for ASD children with happy face stimulus.

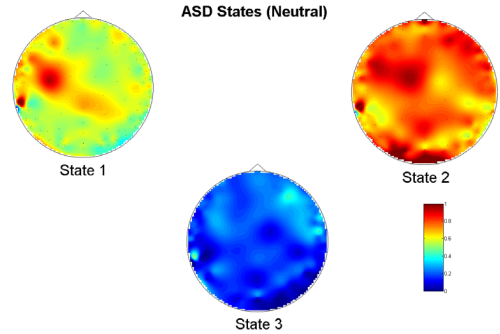

Figure 8. Synchrostates for ASD children with neutral face stimulus.

## B. Brain Connectivity Analysis for Typical and ASD Children

We used the synchronization index (1) to formulate the connectivity graphs for each of the synchrostates corresponding to each of the stimuli with the EEG electrodes representing the nodes and the synchronization index value as the edges between them. Here we only consider those synchrostates which occur the most and the least frequent times (termed as max_state and min_state respectively) during the entire task. The resulting connectivity graphs are shown in Fig. 9-11 with only 4% of the strongest connections retained with the colors representing the degree of synchronization. An interesting observation from Fig. 9–11 is that in general the min_states show more segmented and highly localized connectivity compared to those of the corresponding max_states for all the three stimuli in both the ASD and typical groups. This may mean that most of the specialized information integration operations occur during the min_state and therefore its quantitative characterization may be indicative towards the ability of information integration in ASD and typical children.

TABLE I.    MODULARITY VALUES OF THE MAX/MIN SYNCHROSTATES FOR ASD AND TYPICAL CHILDREN WITH DIFFERENT STIMULUS

| Stimulus | Modularity of max_state | | Modularity of min_state | |
|---|---|---|---|---|
| | *ASD* | *Typical* | *ASD* | *Typical* |
| fear | $1.85 \times 10^{-06}$ | $2.50 \times 10^{-06}$ | $3.72 \times 10^{-06}$ | $1.86 \times 10^{-05}$ |
| happy | $1.82 \times 10^{-06}$ | $2.01 \times 10^{-06}$ | $1.81 \times 10^{-06}$ | $1.86 \times 10^{-05}$ |
| neutral | $1.77 \times 10^{-06}$ | $1.97 \times 10^{-06}$ | $1.96 \times 10^{-06}$ | $1.86 \times 10^{-05}$ |

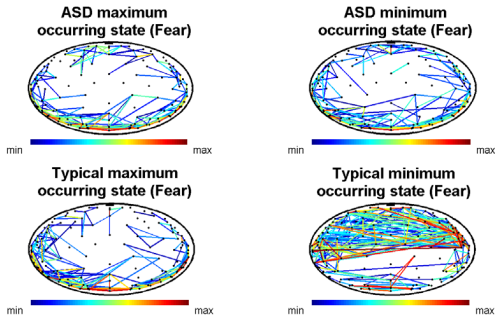

Figure 9. Brain connectivity of typical/ASD with fearful face stimulus.

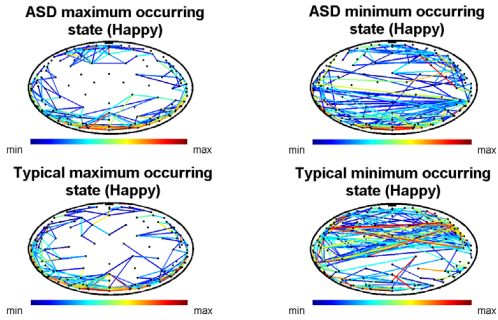

Figure 10. Brain connectivity of typical/ASD with happy face stimulus.

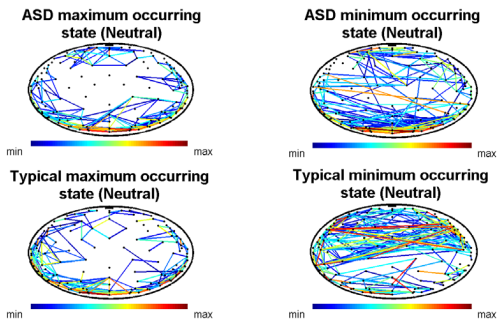

Figure 11. Brain connectivity of typical/ASD with neutral face stimulus.

Table I shows the results of modularity comparison for the two groups under consideration for their respective min_state and max_state. It is evident that for all the stimuli the modularity values of the max_state in both the groups are of the same order whereas the same for the min_state in typical group are consistently an order higher than those in the ASD group. Putting into the perspective of physical meaning of modularity of a network this difference implies that the ASD subjects are less able to do specialized processing during the min_states as their ability to form these localized networks is less than that of the typicals. In one sense this conforms to the findings in the anatomical study of Tommerdahl *et al.* [7]. On the other hand this also shows that modularity could be used as a possible marker for distinguishing ASD from typically-growing children.

## IV. CONCLUSION

We explore the possibility of finding a possible marker to distinguish between ASD and typical population using graph theoretic measure of brain connectivity network. It shows that modularity of the connectivity network formulated following synchrostate analysis of non-invasively recorded EEG data could be an effective identifier of ASD children from age-matched typically-growing ones. Further studies can be performed on individual subjects to explore the degree of distinction modularity values give between these groups.